\documentclass[a4paper]{jpconf}
\usepackage{graphicx}
\usepackage{xspace}
\usepackage{hyperref}
\usepackage{subcaption}

\newcommand{\apenet}{APEnet\xspace}
\newcommand{\apenetp}{APEnet+\xspace}
\newcommand{\PCIe}{PCIe\xspace}

\newcommand{\lofamo}{LO{\textbar}FA{\textbar}MO\xspace}
\newcommand{\quong}{QUonG\xspace}

\newcommand{\eg}{\textit{e.g.}\xspace}
\newcommand{\apelink}{APElink\xspace}

\begin{document}
\title{Architectural improvements and 28~nm FPGA implementation of the
  \apenetp 3D Torus network for hybrid HPC systems}

\author{Roberto Ammendola$^1$, Andrea Biagioni$^2$, Ottorino Frezza$^2$, 
Francesca Lo Cicero$^2$, Pier Stanislao Paolucci$^2$, Alessandro Lonardo$^2$, 
Davide Rossetti$^2$, Francesco Simula$^2$, Laura Tosoratto$^2$, Piero 
Vicini$^2$}

\address{$^1$ INFN Sezione Roma Tor Vergata}

\address{$^2$ INFN Sezione Roma}

\ead{roberto.ammendola@roma2.infn.it}

\begin{abstract}
Modern Graphics Processing Units (GPUs) are now considered
accelerators for general purpose computation. A tight interaction
between the GPU and the interconnection network is the strategy to
express the full potential on capability computing of a
\mbox{multi-GPU} system on large HPC clusters; that is the reason why
an efficient and scalable interconnect is a key technology to finally
deliver GPUs for scientific HPC. In this paper we show the latest
architectural and performance improvement of the \apenetp network
fabric, a \mbox{FPGA-based} \PCIe board with 6 fully bidirectional
\mbox{off-board} links with 34~Gbps of raw bandwidth per direction,
and X8 Gen2 bandwidth towards the host PC. The board implements a
Remote Direct Memory Access (RDMA) protocol that leverages upon
\mbox{peer-to-peer} (P2P) capabilities of Fermi- and
\mbox{Kepler-class} NVIDIA GPUs to obtain real \mbox{zero-copy},
\mbox{low-latency} \mbox{GPU-to-GPU} transfers. Finally, we report on
the development activities for 2013 focusing on the adoption of the
latest generation 28~nm FPGAs and the preliminary tests performed on this new 
platform.
\end{abstract}


\section{Introduction}
We present a status update of \apenetp, which is the high performance,
low latency custom interconnect system developed at INFN targeting
hybrid \mbox{CPU-GPU-based} HPC platforms.
The \apenetp hardware, a \PCIe X8 Gen2 card described
in~\cite{ammendola2012apenet+}, allows building a 3D toroidal mesh
topology of computing nodes.

Moreover, we implemented NVIDIA GPUDirect V1.0 and
V2.0~\cite{GPUdirect} to directly access data on Fermi and Kepler
GPUs.
In this way, real \mbox{zero-copy} \mbox{inter-node}
\mbox{GPU-to-host}, \mbox{host-to-GPU} and \mbox{GPU-to-GPU}
transactions can be issued using a Remote DMA programming paradigm.

At the moment, \apenetp is able to outperform commercial solutions
(like InfiniBand~\cite{Traff:2012:OMB-GPU}) for \mbox{small-to-medium}
message size when using GPU \mbox{peer-to-peer}.
For large message sizes, host memory staging techniques are still
winning, also due to highest bandwidth of latest commercial cards,
which are already \mbox{Gen3-enabled} and guarantee 56~Gbps on the
link.
Of course, our \mbox{mid-term} focus is in upgrading the \apenetp
hardware in order to keep pace with the advances of technology
standards.

\section{Architectural Improvements}

In the \mbox{short-term} period, we focused on the improvement of our
internal architecture.
Three major reworkings have been undertaken regarding the \PCIe
interface, the \mbox{on-board} memory management and the
\mbox{off-board} interface.

\subsection{\PCIe Interface}
On the \PCIe side, we noticed that effective bandwidth on data
transactions was quite low compared to the theoretical one ($\sim 50
\%$).
This is due to the time elapsed between issuing a request on the \PCIe
bus and its completion; this time is system dependent and can be very
large.
In order to optimize the performances (in addition to parameters
tuning like maximum payload size), the card must be able to manage
more than one outstanding request on the \PCIe bus.
In this way, multiple transactions can overlap and total transaction
time is shorter.

At hardware level we needed to implement two concurrent DMA engines
fed by a prefetchable command queue.
The difference between a single DMA and a double DMA implementation
can be seen in Fig.~\ref{fig:2dma}.
We estimated an efficiency gain up of to 40\% in
time~\cite{ammendola:2013:reconfig}.

\subsection{Memory management}
We also focused on removing a known bottleneck on the receiving path,
when virtual to physical address translation is necessary in order to
dispatch data payloads in the correct physical memory areas (whether
they be on host memory or GPU memory).
On \apenetp this task was initially executed by the Nios II embedded
processor but the impact on the resulting execution time was higher
than expected.

\begin{figure}[h]
  \centering
  \begin{minipage}{.46\textwidth}
    \centering
    \includegraphics[width=\textwidth]{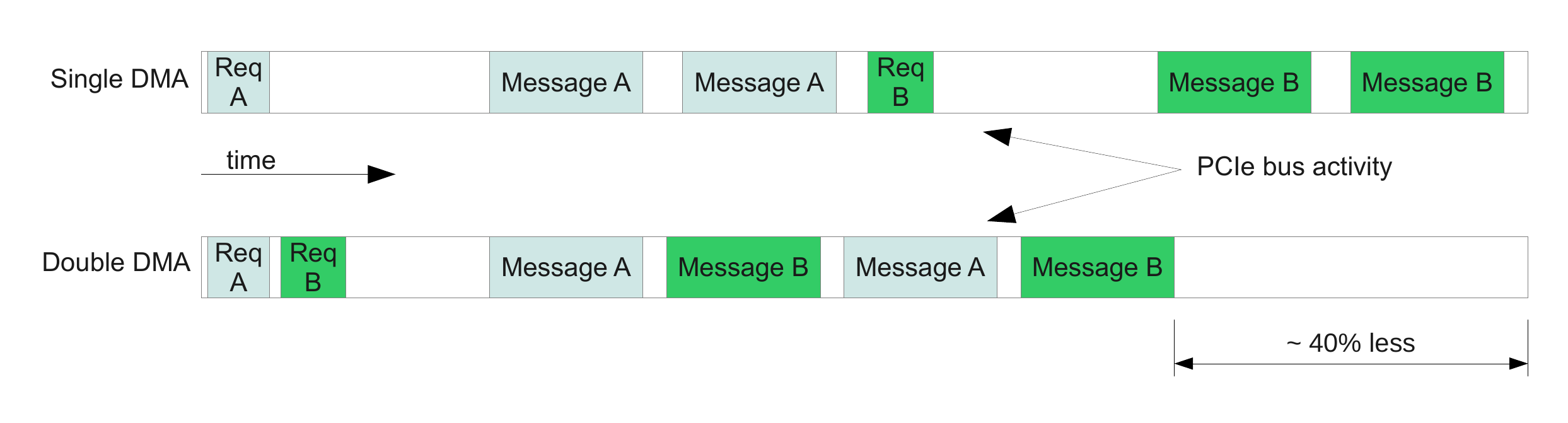}
    \caption{Doubling the number of transaction request on \PCIe bus
      allows an efficiency gain in multiple data transactions (40\%
      reduction in total duration).}
    \label{fig:2dma}
  \end{minipage}
  \hspace{2pc}
  \begin{minipage}{.46\textwidth}
    \centering
    \includegraphics[width=\textwidth]{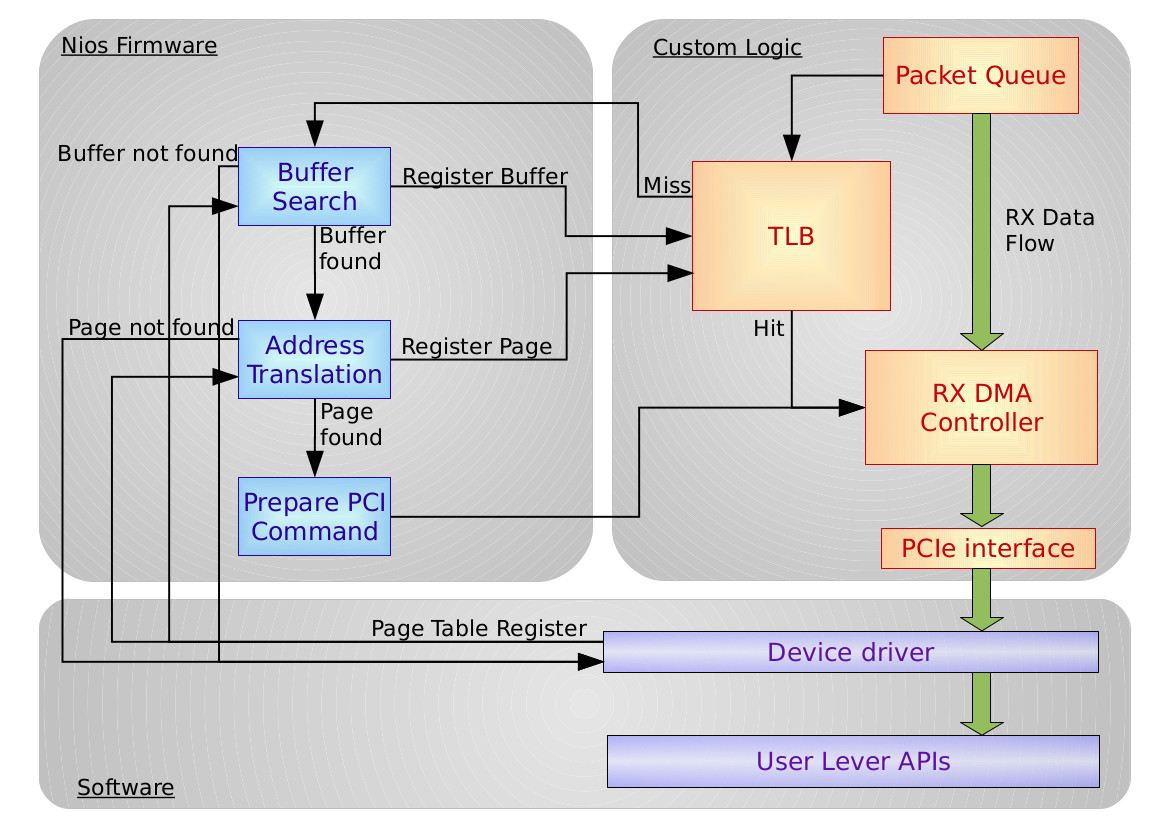}
    \caption{The TLB block performs virtual to physical address
      translation for HW cached pages.}
    \label{fig:tlb}
  \end{minipage}
\end{figure}

Thus, a novel implementation of a Translation \mbox{Look-Aside} Buffer
(TLB) has been developed on the FPGA, to accelerate
\mbox{virtual-to-physical} address translation at hardware
level~\cite{ammendola:2013:FPT}.
As shown in Fig.~\ref{fig:tlb}, the TLB block can store a limited
amount of page entries and, in case of page hit, the Nios II processor
is completely bypassed.
A speedup of up to 60\% in bandwidth on synthetic benchmarks has been
measured with this enhancement.

\subsection{\mbox{Off-board} Interface}
Signal integrity of the transmission system was analyzed in order to
push the embedded transceiver operating frequency to its limits.
Currently, for reliable operations and upper level firmware and
software validation, the Altera transceiver are set at 7.0~Gbps,
yielding a raw aggregated bandwidth of 28~Gbps per \apelink
~\cite{APEnetTwepp:2013} channel.
To estimate the efficiency of the \apelink Transmission Control Logic
operation --- managing the data flow by encapsulating packets into a
light, \mbox{low-level}, \mbox{word-stuffing} protocol --- we devised
a mathematical model; current implementation yields a total efficiency
of $0.784$ over a channel able to sustain $\sim$~2.6~GB/s bandwidth
with a memory footprint limited to $\sim$~40~KB per channel.
%
%
%

\section{Update of Performance Tests}

The described architectural improvements yielded significant
performance gains with respect to our previously published results on
tests of latency and bandwidth.

\begin{figure}[h]
  \centering
  \begin{subfigure}[b]{.32\textwidth}
    \includegraphics[width=\textwidth]{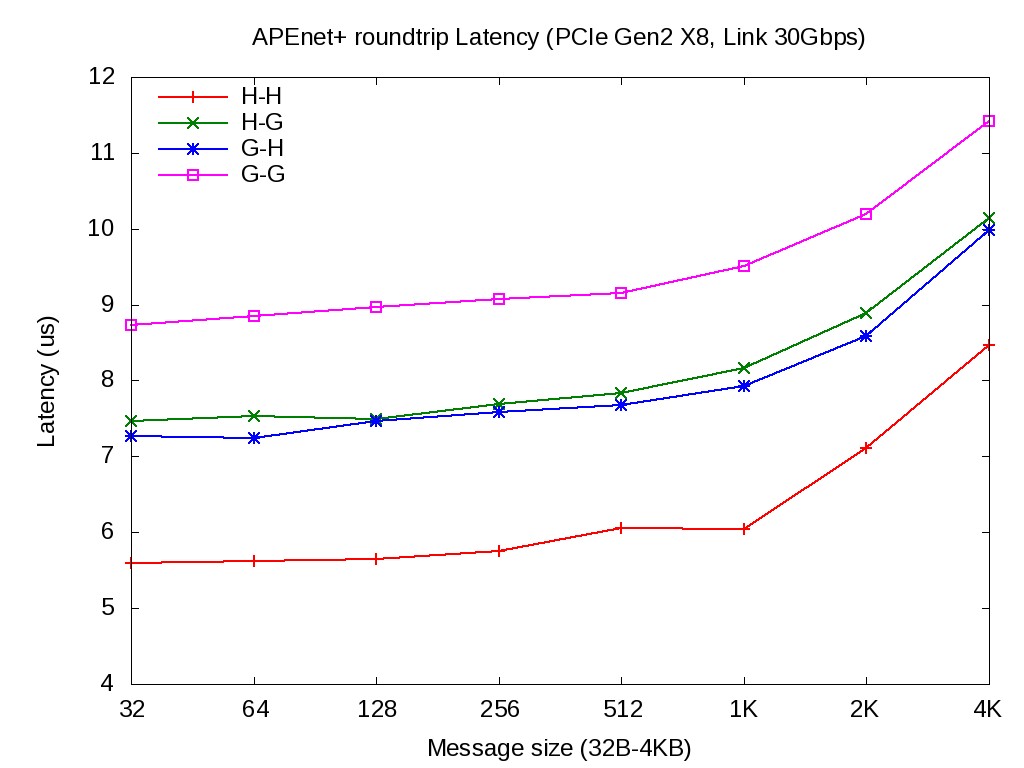}
    \caption{}
    \label{fig:rtlat}
  \end{subfigure}
  \begin{subfigure}[b]{.32\textwidth}
    
\includegraphics[width=\textwidth]{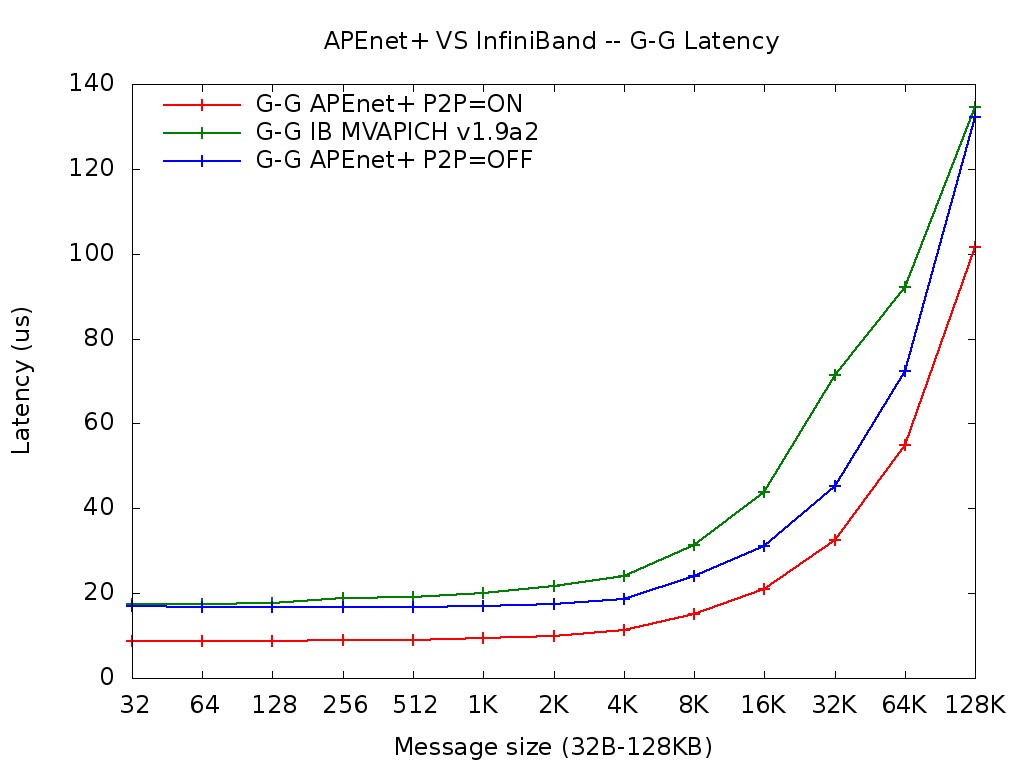}
    \caption{}
    \label{fig:lat}
  \end{subfigure}
  \begin{subfigure}[b]{.32\textwidth}
    \includegraphics[width=\textwidth]{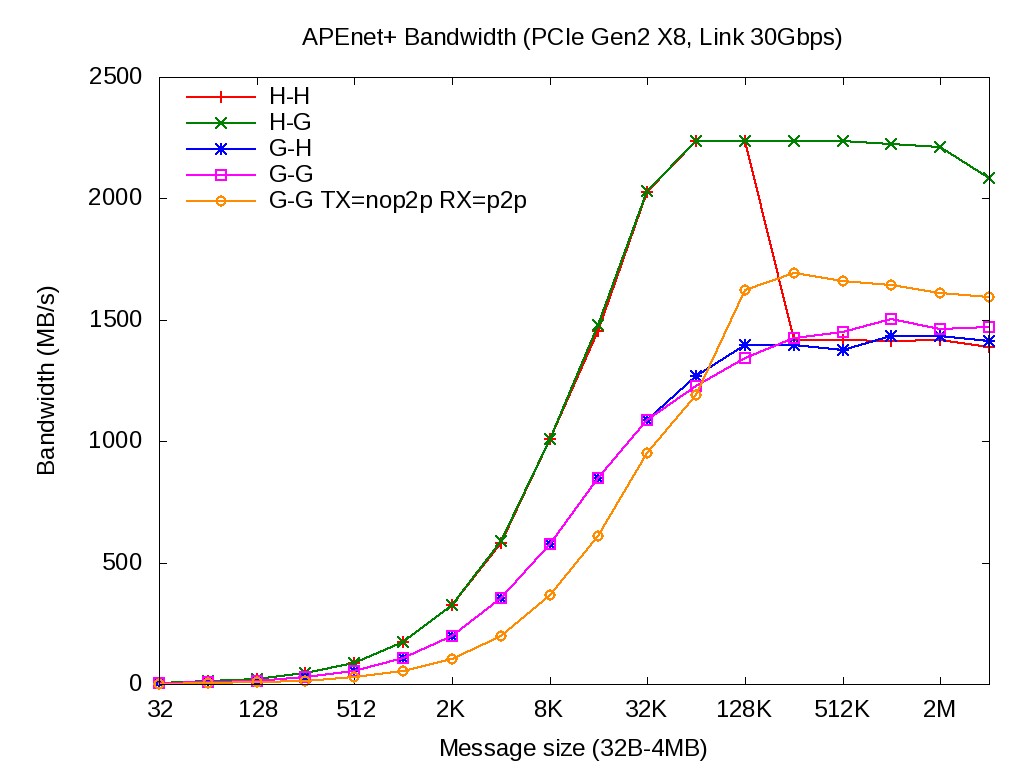}
    \caption{}
    \label{fig:band}
  \end{subfigure}
  \caption{Plots for \mbox{GPU-to-GPU} transfer latency of \apenetp
    vs. InfiniBand with MVAPICH SW stack, Roundtrip latency and
    Bandwidth.}
\end{figure}

On Fig.~\ref{fig:rtlat} we show the round trip latency.
All \mbox{host-bound} and \mbox{GPU-bound} combinations are presented;
the plots clearly show that involvement of the GPU in the transaction
either as sender or receiver causes roughly a 30\% latency increase
for small message sizes.

On Fig.~\ref{fig:lat} we show the advantage of \apenetp P2P technique
over InfiniBand, used with MVAPICH SW stack for message size up to
128~KB; we measure $\sim 8.2 \mu s$ in \mbox{GPU-to-GPU} latency when
using P2P; $\sim 16.8 \mu s$ is measured when P2P is not used; $\sim
17.4 \mu s$ is measured with InfiniBand, on the same platform.

On Fig.~\ref{fig:band} we show the results of several bandwidth tests;
apart from the \mbox{GPU-outbound} cases --- where show that GPU
memory read transactions incur into a bottleneck within the GPU itself
--- in all other transactions (CPU memory read, GPU and CPU memory
write) we can reach the \apenetp link limit, which is $\sim 2.2 GB/s$
on current hardware.



\section{Studies on Fault Awareness}

Fault awareness is the first step when applying Fault Tolerance
techniques in HPC (\eg task migration, checkpoint/restart, \dots).
On the \quong platform, thanks to some \apenetp hardware features,
each node is able to be aware of faults and critical events occurring
to its components and to components of its neighbouring nodes.

Even in case of multiple faults no area of the mesh can be isolated
and no fault can remain undetected at global level.
At the core of this approach, named \lofamo (LOcal FAult MOnitor),
there is a lightweight mutual watchdog protocol between the host node
and \apenetp and the 3D network topology~\cite{LOFAMO2013}.
The \apenetp core contains a \lofamo hardware component and a set of
\lofamo watchdog registers, containing information about the host
status, the \apenetp status and the status of first neighbouring hosts.
On each host in the platform a dedicated \lofamo software component is
able to periodically update the Host Watchdog Register and read the
\apenet Watchdog Register.

In Fig.~\ref{fig:lofamo} it is depicted how a Global Fault Awareness
is obtained, for example, in case a host node stops working; as the
faulty host misses to update its watchdog register, the \apenetp
\lofamo hardware on the same node becomes aware of the fault and sends
diagnostic messages via the 3D network towards its own neighbours.
These hosts can retrieve data about faults occurring on the neighbour
nodes from the watchdog registers and can inform about them (\eg via a
service network) a Master node.
In this way, the Master node has the global picture of the platform
health status and can take decisions about proper countermeasures.

Note that time elapsed since fault occurrence to global fault
awareness is dominated by the watchdog period: for a $WD = 500\,ms$,
$Ta = 0.9\,s$.
In the time range of interest for HPC (watchdog period $1-10^3 ms$),
the addition of \lofamo features has no impact on \apenetp data
transfer latency, as the diagnostic messages are hidden in the
communication protocol.

\section{\apenetp deployment: \quong and other platforms.}

The largest and most significant deployment of \apenetp cards is
within the \quong cluster which is our hybrid, x86\_64 dual GPU
cluster with a $4 \times 4 \times 1$ \apenetp \mbox{3D-torus} network
topology; it is used for testing, development and production run of
scientific codes.
Several \mbox{on-going} projects are developing applications that can
fully exploit our peculiar interconnect solution with promising
results.
Among these, we mention:
\begin{itemize}
\item the simulation of polychronous spiking neural
  networks~\cite{Paolucci:2013:Distributed};
\item a \mbox{Breadth-First-Search} algorithms implementation for
  graph traversal~\cite{Graph400:2012:IA3};
\item a benchmark based on 3D Heisenberg spin glass model by using the
  \mbox{over-relaxation} algorithm~\cite{Bernaschi2013250};
\end{itemize}
Furthermore, in the context of HEP experiments we are testing
designs derived from the \apenetp for \mbox{real-time} GPU
stream processing~\cite{NanetTwepp:2013} and online track
reconstruction with GPUs~\cite{TRIGGU:2012:NSSshort}.


\section{Designing next generation board}

Newer FPGA families --- \eg Altera Stratix V --- are now available on
the market, driving redesign of \apenetp in two major hardware logic
areas: Gen3 migration for \PCIe interface and new transceivers for
increased \mbox{off-board} link speed.

\begin{figure}[h]
\vspace{-60pt}
  \centering
  \begin{minipage}{.46\textwidth}
    \centering
    \includegraphics[width=\textwidth]{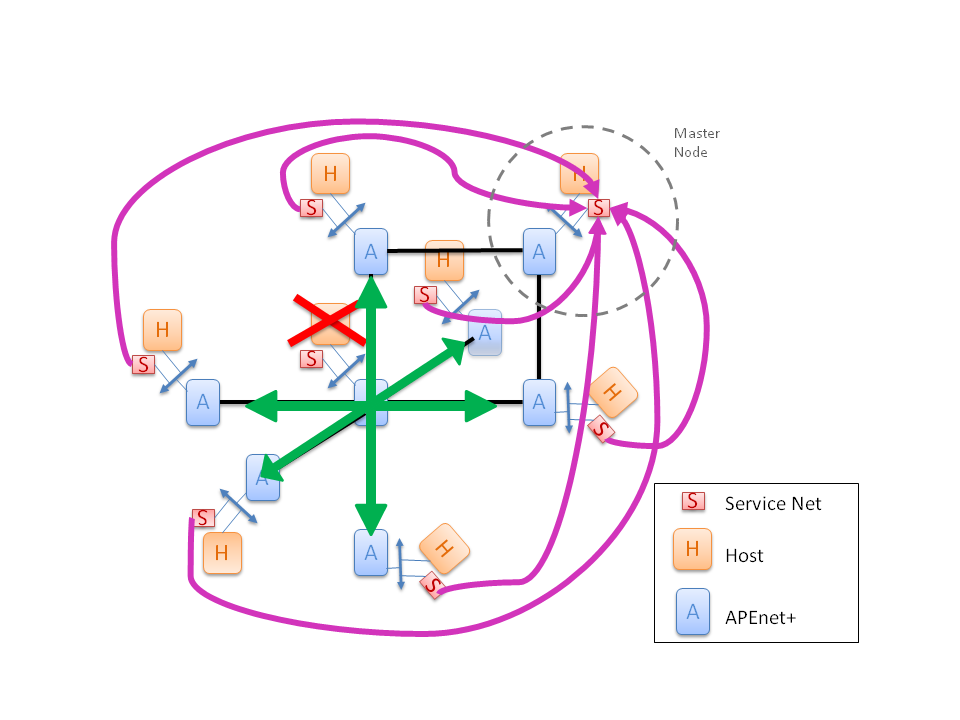}
    \caption{Example of fault detection and awareness at system level with 
LOFAMO.}
    \label{fig:lofamo}
  \end{minipage}
  \hspace{2pc}
  \begin{minipage}{.46\textwidth}
    \centering
    \vspace{10pt}
    \includegraphics[width=\textwidth]{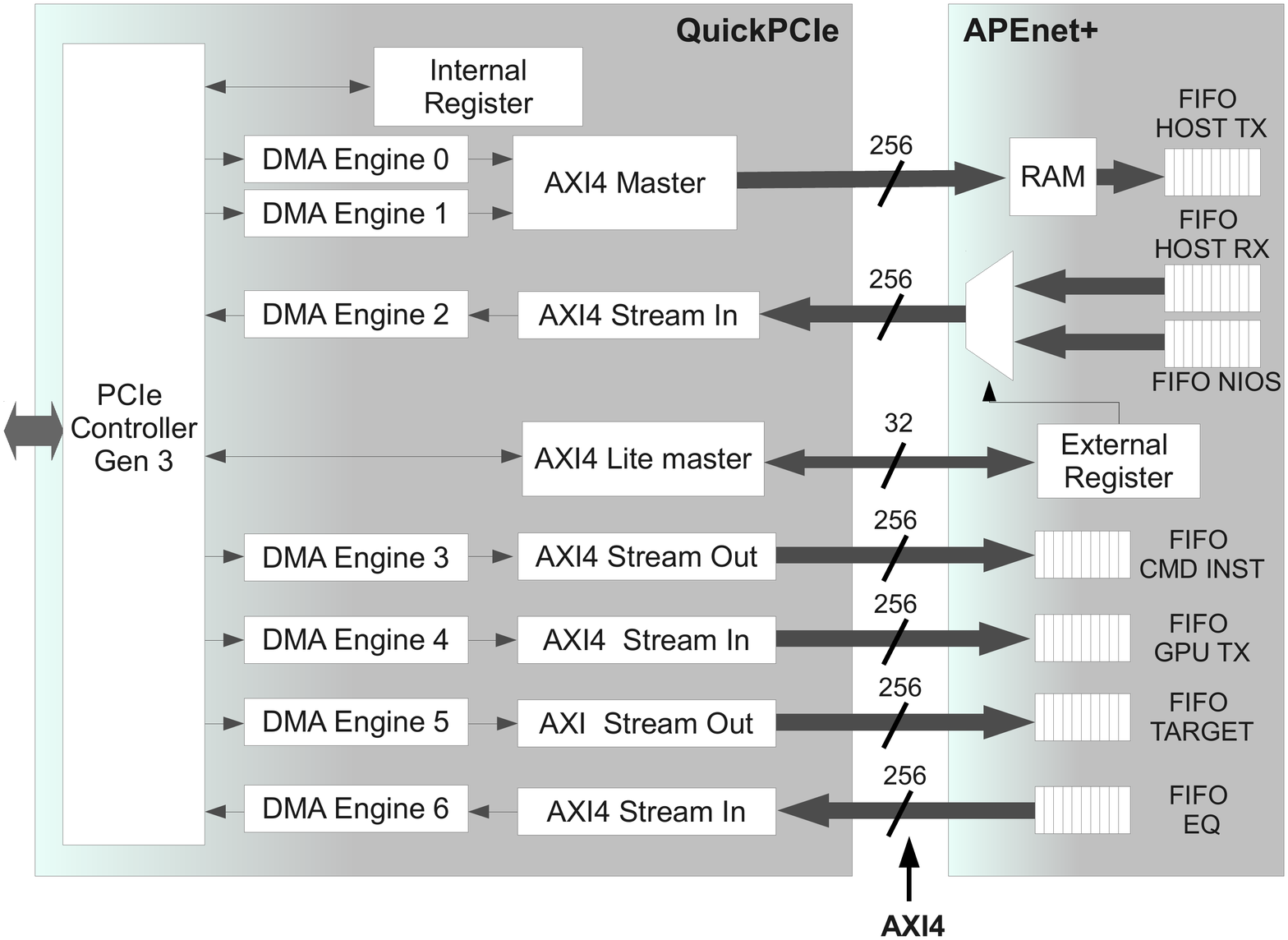}
    \caption{Design of PCIe interface, largely based on AXI4 protocol.}
    \label{fig:pcie}
  \end{minipage}
\end{figure}

\PCIe Gen3 migration allows an increase in bandwidth for the host
interface.
It is based on $8.0 Gbps$ lanes using a 128/130 bit encoding (thus the
protocol overhead is reduced to less than 1\% from 20\% for previous
generations).
The total raw bandwidth that can be obtained with a $\times 8$
interface is $\sim 7.9 GB/s$.
To support this data rate, on the back-end the data-path must be
\mbox{256-bit} wide, with a clock reference of 250 MHz.
The standard used is AXI4~\cite{alteraqsys}, which needed a redesign
of \apenetp internal \PCIe interface, as depicted in
Fig.~\ref{fig:pcie}.
AXI4 migration is also preparatory for future use of embedded ARM hard
IP processors, foreseen on \mbox{high-end} class FPGAs only like the
future Stratix 10 devices.

New Altera devices are capable of 14.1~Gb/s transceivers, which can be
bonded in 4 lanes to build up a 56~Gb/s \mbox{off-board} link.
As a physical medium we can rely on QSFP+ standard that has been
upgraded to work at these data rates (the same as InfiniBand FDR).

In order to develop \PCIe Gen3 migration and 56~Gb/s class
\mbox{off-board} links, we used an Altera development board with a
Stratix V GX FPGA \cite{alteraS5devboard}.
We implemented the link using the single 40~Gb/s QSFP+ onboard
connector and performed data transfer tests between 2 such boards.
As a preliminary result we achieved a link speed of 11.3~Gbps/lane
(45.2~Gbps/channel), still using \mbox{40~Gb/s-certified} cables.

\section{Conclusion}

We presented a status update of the development of our custom interconnect 
system \apenetp. Several architectural improvements have been discussed, that 
brought to substantial performance enhancements. We also introduced the
fault-aware capability now embedded in the \apenetp communication protocol, 
which can be used in advanced high-level fault tolerance techniques. We 
finally reported on latest developments on 28 nm FPGA devices, that allow us to 
upgrade the \PCIe interface to Gen3, and the Off-board link to 56 Gb/s data rate.

\section*{Acknowledgement}
This work was partially supported by the EU Framework Programme 7 project 
EURETILE under grant number 247846; Roberto Ammendola was supported by MIUR
(Italy) through INFN SUMA project.

\section*{References}

\providecommand{\newblock}{}


\begin{thebibliography}{10}
\expandafter\ifx\csname url\endcsname\relax
  \def\url#1{{\tt #1}}\fi
\expandafter\ifx\csname urlprefix\endcsname\relax\def\urlprefix{URL }\fi
\providecommand{\eprint}[2][]{\url{#2}}

\bibitem{ammendola2012apenet+}
Ammendola R {\em et~al.\/} 2012 {\em Journal of Physics: Conference Series\/}
  {\bf 396} 042059
  \urlprefix\url{http://stacks.iop.org/1742-6596/396/i=4/a=042059}

\bibitem{GPUdirect}
{NVIDIA} {GPUD}irect technology \url{https://developer.nvidia.com/gpudirect}

\bibitem{Traff:2012:OMB-GPU}
Bureddy D, Wang H, Venkatesh A, Potluri S and Panda D 2012 {\em Recent Advances
  in the Message Passing Interface\/} ({\em Lecture Notes in Computer
  Science\/} vol 7490) ed Tr\"aff J~L, Benkner S and Dongarra J~J (Springer
  Berlin Heidelberg) pp 110--120 ISBN 978-3-642-33517-4
  \url{http://dx.doi.org/10.1007/978-3-642-33518-1_16}

\bibitem{ammendola:2013:reconfig}
Ammendola R {\em et~al.\/} 2013 {\em Track on Interconnect architectures for
  reconfigurable computing systems, held at Reconfigurable Computing and
  {FPGA}s (ReConFig), 2013 International Conference on\/} to be published

\bibitem{ammendola:2013:FPT}
Ammendola R {\em et~al.\/} 2013 {\em Field-Programmable Technology (FPT), 2013
  International Conference on\/} to be published

\bibitem{APEnetTwepp:2013}
Ammendola R {\em et~al.\/} 2013 {\em JINST, Journal of Instrumentation,
  Proceedings of Topical Workshop on Electronics for Particle Physics (TWEPP)
  2013\/} (IOP Publishing) to be published

\bibitem{LOFAMO2013}
Ammendola R {\em et~al.\/} 2013 {\em arXiv:1307.0433\/}
  \url{http://arxiv.org/abs/1307.0433}

\bibitem{Paolucci:2013:Distributed}
Paolucci P {\em et~al.\/} 2013 {\em arXiv:1310.8478\/}
  \url{http://arxiv.org/abs/1310.8478}

\bibitem{Graph400:2012:IA3}
Bisson M, Bernaschi M, Mastrostefano E and Rossetti D Breadth first search on
  \apenetp \url{http://cass-mt.pnnl.gov/docs/Session 2 - 1.pdf} iA3 Workshop on
  Irregular Applications: Architectures and Algorithms, in conjunction with
  Super Computing 2012

\bibitem{Bernaschi2013250}
Bernaschi M, Bisson M and Rossetti D 2013 {\em Journal of Parallel and
  Distributed Computing\/} {\bf 73} 250 -- 255 ISSN 0743-7315
  \url{http://www.sciencedirect.com/science/article/pii/S0743731512002213}

\bibitem{NanetTwepp:2013}
Lonardo A {\em et~al.\/} 2013 {\em JINST, Journal of Instrumentation,
  Proceedings of Topical Workshop on Electronics for Particle Physics (TWEPP)
  2013\/} (IOP Publishing) to be published

\bibitem{TRIGGU:2012:NSSshort}
Amerio S {\em et~al.\/} 2012 {\em Nuclear Science Symposium and Medical Imaging
  Conference (NSS/MIC), 2012 IEEE\/} pp 1806--1811 ISSN 1082-3654

\bibitem{alteraqsys}
Altera {Q}sys {I}nterconnect, {Q}uartus {II} 13.0 {H}andbook, {V}olume 1
  \url{http://www.altera.com/literature/hb/qts/qsys_interconnect.pdf}

\bibitem{alteraS5devboard}
Altera {S}tratix {V} {GX} {D}evelopment {B}oard
  \url{http://www.altera.com/products/devkits/altera/kit-sv-gx-host.html}

\end{thebibliography}

\end{document}